# The 1/4 occupied O atoms induced ultraflat band and the one dimensional channels in the Pb10−xCux(PO4)6O4 (x=0,0.5) crystal

Kun Tao,[1] Rongrong Chen,[1] Lei Yang,[1] Jin Gao,[1] Desheng Xue,[1] and Chenglong Jia[1,2]

[1] Key Lab for Magnetism and Magnetic Materials of Ministry of Education, Lanzhou University, Lanzhou, China

[2] Lanzhou Center for Theoretical Physics and Key Laboratory of Theoretical Physics of Gansu Province, Lanzhou University, Lanzhou, China

The search for room-temperature superconductors has been a long-standing goal in condensed matter physics. In this study, we investigate the electronic and geometric properties of lead apatite with and without Cu doped within the frame work of the density functional theory. Based on our calculations, we found that without the Cu doped the lead apatite shows an insulator character with flat bands straddle the Fermi level. Once we introduce the O1 vacancies, the flat bands disappear. Furthermore, we analyze the effects of Cu doping on the crystal structure and electronic band structure of the material. Our calculations reveal the presence of one-dimensional channels induced by fully occupied O1 atoms, that are only 1/4 occupied in the literature, which may play a crucial role in the realization of room-temperature superconductivity. Based on our findings, we propose a possible solution to improve the quality of superconductivity by annealing the material in an oxygen atmosphere. These results contribute to a better understanding of the unusual properties of Cu-doped lead apatite and will pave the way for further exploration of its potential as a room-temperature superconductor.

Being the holy grail of condensed matter physics, the high-TC superconductors due to their unique characteristics of zero resistance and perfect diamagnetism [1] with huge potential applications for an energy-efficient future and are used in a billion-dollars market. Milestones of superconductivity research include its discovery by Onnes in 1911, the BCS theory of Bardeen, Cooper and Schrieffer [2], and the discovery of highCTc superconductors by Bednorz and Müller [3], and more recently that of hydride superconductors by Eremets and coworkers [4] and the subsequently followed by several new classes including the Fe-pnictides in 2008 [5] and as well as nickelate superconductivity by Li et al. [6] Moreover, twisted bilayer graphene [7, 8] with the flat bands near the Fermi level, which is a signal of the superconductivity have been discovered. Strong correlation bands are a common feature of many high TC superconducting families, which will generate unconventional mechanisms for Cooper pair formation. Strong correlation bands are a common feature of many high TC superconducting families, which will bring about new unconventional mechanisms that arise.

Finding a room-temperature superconductor has always been the ultimate dream for the condensed matter physicist. However, a definitive road-map to achieving room temperature TC under ambient pressures is still remained elusive. While the recently claimed in a nitrogen-doped lutetium hydride [9] at near-ambient pressure superconductivity still remains a subject of ongoing debate [10–13]. However, it was recently reported by Lee et al. [14, 15] who proclaim that they have successfully synthesized even at the ambient pressure such a room temperature superconductor, Pb10−xCux(PO4)6O with $0.9 < x < 1$. The recipe to synthesize Pb10−xCux(PO4)6O appears quite easy for other groups are also easy to follow up with[16–18]. This announcement has caused a carnival within the physics community and beyond[19, 20]. In the wake of this claim, numerous independent groups have tried to duplicate the synthesis of LK-99 and its claimed superconductivity, and have produced a variety of results[21–26]. Several experimental group have confirmed the absence of electronic resistance in LK-99 below 100 K at ambient pressure, though they didn't observe the Meissner effect[21]. Others have detected pronounced diamagnetism without resistivity measurement[22]. On the contrary, several studies have not found any indications of superconductivity or diamagnetism, identifying LK-99's behavior as either semiconducting, insulating, or paramagnetic[23–26]. Several theoretical analyses have explored its electronic properties based on the density functional theory. Lead apatite was claimed to exhibits insulating behavior from the experimental groups, but contrary to this, DFT calculations show that the Cu doping introduces correlated flat electronic bands of Cu-d character at the Fermi level[27–37]. This is an iconic feature often seen in high-TC superconductors. Despite these efforts, the questions about the potential superconductivity of Cu-doped lead apatite still persist, emphasizing the urgent requirement for further exploration into its crystal structure, electronic structure, and other intrinsic properties[38–40]. Nevertheless, it is exciting and a definitely call for a more thorough theoretical understanding of this rather unusual material.

In the paper, we performed the first principle calculations to investigate the geometrical and the electronic structure of the lead apatite. With carefully analyzing of the geometry and the electronic structure of the before and after the Cu doped lead apatite, we emphasize that the quality of the superconductivity in LK-99 is determined by the presence of the p states of the 1/4 occupied O1 atoms that are fully occupied. Based on our calculations, we proposed a possible solution method to improve the quality of the LK-99.

Our calculations are performed in the framework of Density Functional Theory (DFT) as implemented in the Vienna Ab initio Simulation Package (VASP) package [41, 42] with the projector augmented wave (PAW) potentials. The basis set contained plane waves with a kinetic energy cutoff of 520 eV and the total energy was converged to $10^{-6}$ eV. The Brillouin zone integration is carried out with 4×4×4 k-pointing sampling for undoped structures and with a Gamma-pointing sampling for the Cu doped structures of a 2×2 supercell of the stoichiometry of the Pb10−xCux(PO4)6O crystal and the

Pb10−xCux(PO4)6O4 with x=0, 0.5. All geometries, including the lattice constants, the cell shape and the cell volume, were optimized without any symmetry constraint until all residual forces on each atom were less than 0.01 eV/A.

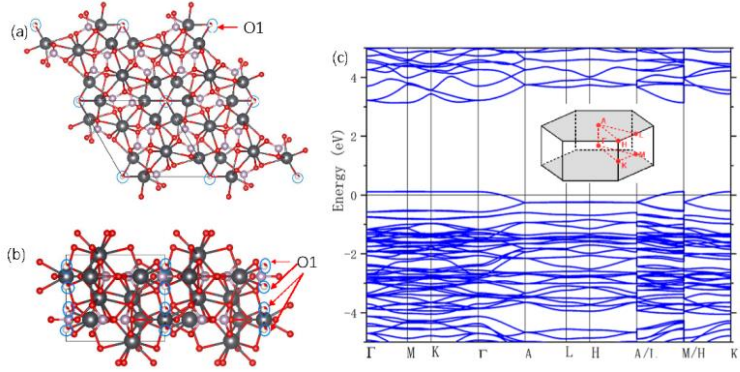

Fig.1 The top view (a) and the side view and (b) of the crystal structure of lead-apatite with a chemical formula (Pb10(PO4)6O4) that used in our calculations. respectively. The atoms circled in blue are the O1 atoms. (c) The calculated band structure of the lead apatite. The inset in (b) shows the high symmetry k-points selected for plotting the band structure in (a)

The structure of the Pb10(PO4)6O4 is presented in Fig.1. The structure possesses P63/m symmetry (space group no. 176) with a hexagonal lattice. Here, we consider the lead apatite. Taking its structure reported from X-ray diffraction in Ref. [43] and from COD (Crystallography Open Database) data of the pure lead-apatite as the parent compound. We noticed that in the lead-apatite, there exists two symmetry inequivalent Pb atoms, named Pb1 and Pb2, and the Pb1 atoms form a hexagon, while the Pb2 atoms form two oppositely shaped triangles. The O1 atoms are only 1/4 occupied, as it is not part of the six PO4 tetrahedron, which are marked with the blue circles, as shown in the Fig.1 (a), (b). The other O2 atoms which form the six PO4 tetrahedrons. We should point out that these two types of the O atoms are inequivalent. As for our calculations, we used a fully occupied O1 atoms which are only 1/4 occupied in the literature[27−37], in which three of the four O1 are removed to keep the stoichiometry as Pb10(PO4)6O, that is the reason why our formula of the lead apatite (Pb10(PO4)6O4) is different with the stoichiometric formula Pb10(PO4)6O in the reported papers. Here, we want to emphasize the importance of the fully occupied O1 atoms, which are only 1/4 occupied in the literature. As will be discussed in the following parts, this is a key parameter for the possible room temperature superconductor.

The lattice constants are optimized with the VASP code with the ionic positions, cell volume, and cell shape are all allowed to change. The optimized lattice are a=b=10.165 ˚A and c=7.367 ˚A, in good agreement with the experimental values a=b=9.865 ˚A and c=7.431 ˚A. The predicted larger lattice parameters may arise from the overestimation of the employed PBE functional.

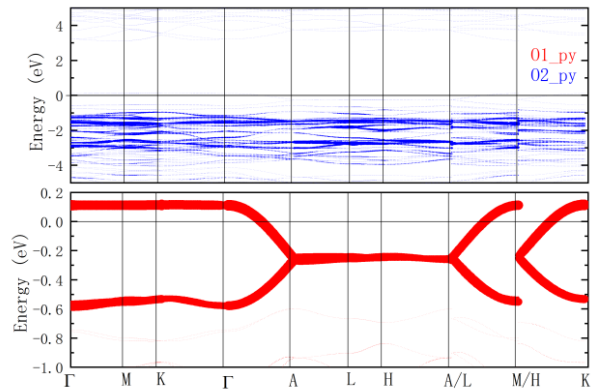

Fig.2 (a) The projected band structure for the O1 and the O2 atoms. (c) The projected density-of-states for the O1 and the O2 atoms, respectively.

The calculated band structure of the lead apatite is shown in the Fig.1 (c). It can be observed from the Fig.1(c) that there exist an ultraflat band along the ΓM-K-Γ and also A-L-H-A path, and the corresponding band width is near 0.24 mev and the 2 meV. And there is also an indirect band gap of about 2.72 eV between the Γ point and the middle point along the

Γ-M path. The obtained insulating nature of lead apatite is consistent with the experimental observations and other DFT calculations.[14, 15].

There is a peak just locates at the Fermi level which is mainly caused by the hybridization between the O py(px) and the Pb s orbitals, as shown in the supplementary Fig.S1, which results in an ultraflat band just about 0.126 eV above the Fermi level. Full occupied O1 atoms are very important, which results in a metal, as shown in the Fig.2 (b). With a further investigation on the two types of the O atoms, it can be found from the Fig.2 that the O1 atoms give the vital contributions to the flat band. It means that the fully occupied O1 atoms reported in the literature determines the flat band. Or in other words, since the O1 atoms are only 1/4 occupied, the O1 vacancies determines whether the flat bands exist or not.

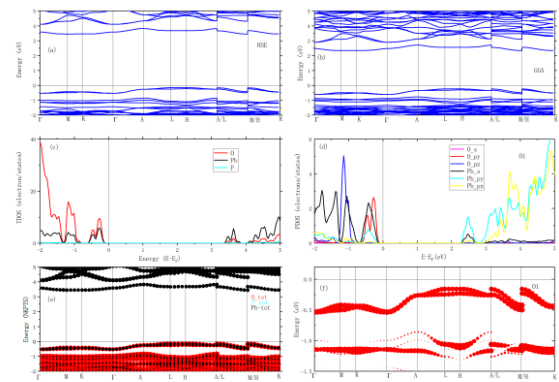

Fig.3 (a),(b) The calculated band structure with the GGA and the HSE methods for the Pb10(PO4)6O crystal with the O1 vacancy after the fully relaxations. (c) The total DOS for the O, Pb, and P atoms. (d) The pdos for the O and the Pb atoms. (e) The total band structure for the O, Pb and P atoms. The projected band structure on the O1.

In order to check our assumption right or not, we removed three out of four O1 atoms in calculations forming the stoichiometry formula Pb10(PO4)6O structure as reported by the experiments[14, 15] and other DFT calculations[27–40]. The band structure is plotted in the Fig.3(a). As we know the DFT calculations can not reproduce accurate band gaps, thus the Heyd-Scuseri-Ernzerhof (HSE06) functional is employed for the band structure calculations[46]. From the Fig.3(a) and (b), it can be found that the two band structures are very similar, yet have different band gaps. The band gap increase from 2.72 eV in the GGA approximation to 3.61 eV with the HSE method. It can also be observed that the flat bands in the stoichiometry formula Pb10(PO4)6O structure disappear. Because the bond length between the Pb1 atoms with that of the O1 atoms change from the 2.58 ˚A to 2.27 ˚A, such strong local distortions will result in the disappear of the flat band. A nearly flat band along the A-L-H-A path with the band width of about 0.11 eV can be observed but below the Fermi level about 0.25 eV, as shown in the Fig.3 (a). It is due to the hybridization between the Pb s and px(py) orbitals and the O py (px) orbitals. With a further observation of the pbands for the O1 atoms, as shown in the Fig.3 (f), it can be found that the flat band is mainly composed by the O1 py (px) orbitals. Therefore, we may conclude that the fully occupied O1 atoms determine the electronic structure of the undoped lead-apatite crystal.

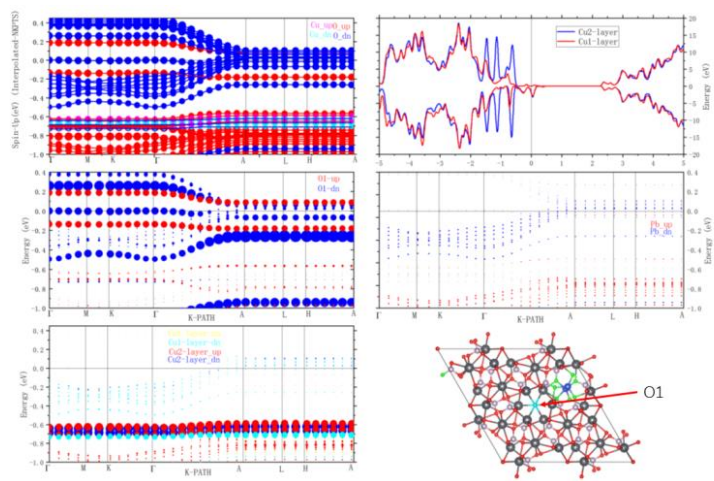

Fig.4 (a) The calculated band structure of the LK-99-O4 crystal. (b)The layer dependent DOS of the LK-99-O4 crystal. (c) and (d) The projected band structure on the O and the Pb atoms. (e) The layer dependent pband of the LK-99-O4

crystal.(f) The top view of the structure of the LK-99-O4 crystal.

After analyzing structural and electronic properties of lead-apatite, we turn to the Cu doped lead-apatite which has a formula Pb9.5Cu0.5(PO4)6O4 (noted as LK-99-O4) in our calculations. Flat bands play a significant role in determining the physical properties of materials, including their potential for superconductivity. The quantum geometry of flat bands is crucial in determining the nature of the resulting strongly correlated phases. For instance, in the case of repulsive interactions, flat bands can lead to ferromagnetism, while attractive interactions can potentially give rise to superconductivity. Therefore, we perform the spin polarized calculations with taking into account the on-site Coulomb interactions Ueff =4.0 eV for the Cu-3d shell. The LK-99 crystala in experiment and previous calculations has a chemical formula of Pb10−xCux(PO4)6O with 0.9 <x<1.1 [14, 15, 28], that is the reason why our results differ from previous reported DFT calculations results. Furthermore, it was reported that the Cu atoms will occupy the Pb2 positions in the LK-99 crystal [14, 15]. In order to simulate the Cu substitute the Pb2 site, we use a 2×2 supercell with two Pb2 atoms are replaced by two Cu atoms, resulting in LK-99O4 crystal. There are several possible configurations for the Pb atoms to be replaced by the Cu atoms. After fully relaxations, we found an energy more favorable configuration as as shown in the Fig.4.(f). The lattice constant of our LK-99-O4 is a= 19.676 ˚A. b= 19.721 ˚and c= 7.260˚ A, and the two Cu atoms are nearly at the same position, eg Pb2 position, but one Cu atom (noted as Cu1) is about 2.9 ˚higher than the other Cu atom (noted as A Cu2).

Comparing with the Pb10(PO4)6O crystal, it can be found that the doped Cu in the LK-99-O4 crystal induced some extraordinary ultraflat bands along the Γ-M-K-Γ and along the A-L-H-A path just straddle the fermi level, as shown in the Fig.4 (a). In order to further check the effect of the Cu atoms on the superconductivity, we also plot the band projected on to the layer where Cu1 and Cu2 atoms are located. It can be observed that they both contribute minimally to the flat band that straddles the Fermi level, please see the supplementary material Fig.S2 for further details. However, they significantly contribute to the layer decomposed band projection. It can also be observed from the Fig.4 (f) that the Cu doped induced some local lattice distortions. The bond length between the Pb1 and the O1 atoms changes from the 2.70 ˚A to the 2.84 ˚A, such a large increase in the Pb-O bond length will induce a strong local lattice distortion, which leads to the splitting of the O1 p orbitals into the px (py) and the pz orbitals, as shown in the supplementary Fig.S1. In comparison with the undoped lead-apatite, Cu doping results in an insulator-metal transition which was claimed as one of the key factors contributing to the realization of room-temperature superconductivity in LK-99 [14, 15].

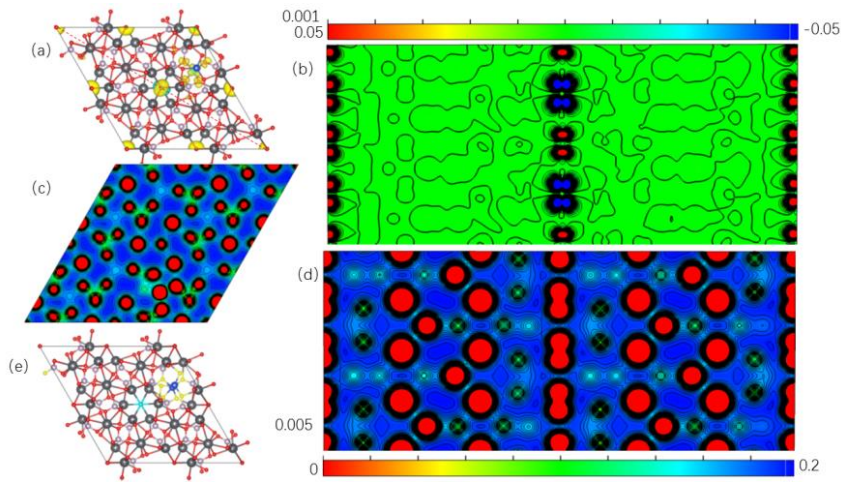

Fig.5 (a), (c)The magnetization charge density of the LK-99-O4 crystal. (b) The slice along the long diagonal in (a). (c) The slice of the total charge density for the Cu2-layer. (d) The slice of the total charge density along the long diagonal of the LK-99-O4 crystal.

The doped Cu induced ultraflat band is mainly due to the hybridizations between the spin down px (py) orbial of the O1 atoms and the spin down s orbitals of the P-b atoms, as it can also be observed from the Fig.4 (c) and (d). The flat bands are usually considered due to the hybridizations beteen the Cu d orbital and its near O orbital[47]. However, according to our spin polarized calculations, the energy bands of the two Cu atoms primarily lie 0.6 eV and 0.8 eV below the Fermi level, resulting in magnetic moments of 0.672 μB and 0.03 μB for Cu2 and Cu1, respectively. With a carefully analyze of the PDOS of the O1 atoms and the Pb atoms, as illustrated in the Fig.S1, can be observed that the induced flat bands primarily result from the hybridizations between the O1 px (py) and Pb s orbitals, straddling the Fermi level. This leads to magnetic moments for the O1 and Pb atoms ranging from -0.268 μB to 0.455 μB and from -0.002 μB to 0.012 μB respectively, resulting in a total magnetic moment of approximately 8.002 μB for the supercell. This is demonstrated in the magnetic charge density for the supercell as shown in the Fig.5 (a). According to Fig.5 (b), it can be observed that there exist two types of one-dimensional channels. One is due to the periodically changed spin up and spin down of the center O1 atoms, while the other one is composed by the spin up part of the corner O1 atoms. Both of these "grilled skewers" like one dimensional channels are encircled by their neighboring Pb atoms, which generate an essential electrostatic field. However, due to their high atomic mass, lead cations have reduced mobility and do not block the channels even at ambient temperature, which may be the key factor contributing to the occurrence of superconductivity.

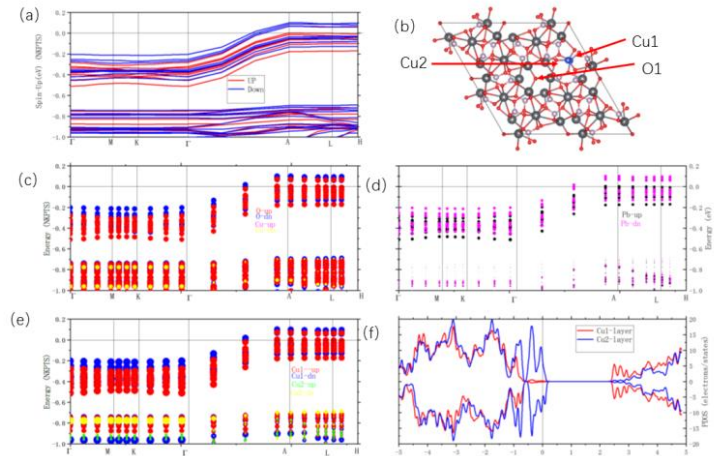

Fig.6 (a) The calculated band structure of the LK-99-O1 crystal. (b)The top view of the structure of the LK-99-O4 crystal. (c) and (d) The projected band structure on the O and the Pb atoms. (e)and (f) The layer dependent pband and PDOS of the LK-99-O1 crystal.

Since there are two types of the O atoms O1 and the O2, as shown with different color in the Fig.5 (e), it should give further investigations on the contribution of each O atoms to the flat band. From the Fig.S3, it can be observed that there is a big difference in the contribution to the flat bands for the two types of the O atoms. The fully occupied of the 1/4 occupied O1 atoms give the vital contributions to the flat band, while that of the O2 atoms only give very small contributions to the flat band. Based on the above analysis, we can obtain a conclusion that the O1 vacancy will have a vital effect on the electronic structure of the LK-99-O4 crystal. Moreover, here a new model should be used to describe the physics of LK-99-O4, unlike the widely studied one-band Hubbard model applicable to cuprates [44, 45], since the bands are contributed by the hybridization between the Pb s states with its next next nearest-neighboring O1 atoms py(px) states. In comparison with the non-doped lead-apatite, Cu doping creates two one-dimensional channels, which may be one of the crucial factors contributing to the realization of room-temperature superconductivity in LK-99-O4.

To further reveal the effect of the O1 atoms on the superconductivity, we also performed the calculations with the Cu doped into the stoichiometry formal Pb10(PO4)6O as reported above, forming the Pb9.5Cu0.5(PO4)6O (noted as LK-99-O1) crystal, as shown in the Fig.6 (a). After fully relaxations, we found an energy more preferable configuration as shown in the Fig.6 (a) with the optimized lattice constant is a= 19.677 Å, and C= 7.289 Å, b= 19.575 Å. In this structure, the Cu1 atom is nearly just above the Cu2 atom, but with about 3.64 Åhigher than that of the Cu2 atom. Here, we should emphasize that the Cu1 atom and one of the O1 atom are in the same plane, while in the LK-99O4 crystal neither of the Cu1 and the Cu2 atoms are in the same plane with any of the O1 atoms. This difference will lead to a significant difference in the layer-dependent contribution to the potential superconductivity, as will discussed in the following.

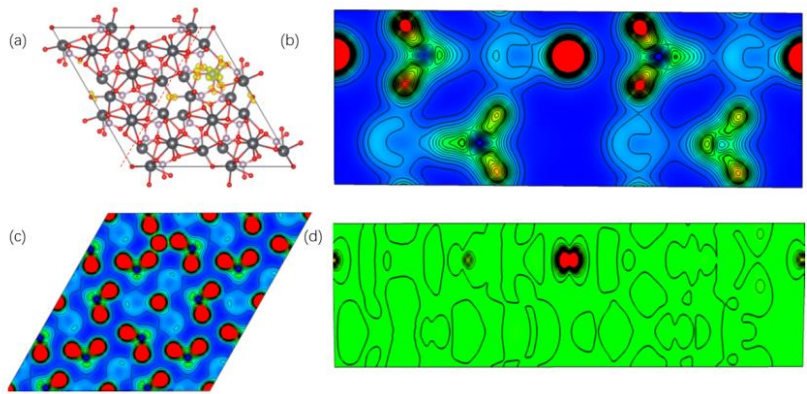

Fig.7 (a), (c)The magnetization charge density of the LK-99-O1 crystal. (b) The slice along the short diagonal in (a). (c)

The slice of the total charge density for the Cu1-layer. (d) The slice of the magnetization charge density along the short diagonal of the LK-99-O1 crystal.

Similar to that of the LK-99-O4 crystal, the doped Cu in the LK-99-O1 crystal induces some extraordinary flat bands. However, these flat bands are only along A-L-H kpath and straddle over the Fermi level. Moreover, it can be observed that there is a band gap about 2.65 eV between the HOMO and the LUMO along the Γ-M-K-Γ path. These flat bands are induced by the hybridization between the spin up of Pb orbital and the spin up O1 orbitals, as shown in the supplementary Fig.S4, while in the LK-99-O4 crystal, these flat bands are induced by the hybridization between the spin down of Pb orbital and the spin down O orbitals. Similarly, the pbands of the two Cu atoms are also about 0.6 eV to 1.0 eV far below the Fermi level, which leading to the magnetic moment of 0.601 μB and 0.496 μB for the Cu1 and Cu2 and about 1.894 μB for the supercell, as shown in the magnetic charge density for in the Fig.7 (a). Furthermore, we examine the slice of magnetic charge density across the supercell's short diagonal and discover that the presence of one-dimensional channels in the LK-99-O4 crystal disappears. The disappearance is attributed to the fact that in the stoichiometry formal LK-99-O1 crystal, three out of four O1 atoms are deleted, leaving only one O1 atom. This could be the reason why the superconductivity is not observed in the experiments. To examine the influence of Cu on the potential superconductivity further, we have plotted the layer dependent band projected onto the Cu atoms as in Fig.7 (f). It can be observed that a flat band straddling the Fermi level only exists in the Cu1 layer.

It is reported that the LK-99 sample could potentially exhibit multiphase properties. One group claimed that the crystals are highly insulating and optically transparent[48], which consistent with our calculations results for the LK-99-O1. The other group's conclusion are that the material could more likely be a magnet[49], which is coincident with our calculations results for the LK-99-O4. The presence of O1 vacancies or the absence of an optimal oxygen atmosphere during sample preparation may affect the formation of superconducting states and hinder the observation of superconductivity. By the way, the authors claimed the superconductivity may be due to two factors[14, 15]: the first factor is the volume contraction resulting from an insulator-metal transition achieved by substituting Pb with Cu. We indeed find that the insulator-metal transition by replacing the Pb atoms with the Cu atoms. And the second factor contributing to the realization of superconductivity at Tc is the increased on-site repulsive Coulomb interaction, which is enhanced by the structural deformation in the one-dimensional (1D) chain (Pb2-O1/2-Pb2 along the c-axis) structure due to the formation of superconducting condensation. However, according to our results we found the O1 induced two kinds of the one dimensional channels, which maybe the key factor for the superconductivity.

Based on the above discussions, we may conclude that the hybridization between the Pb s-states with the p state of the fully occupied 1/4 occupied O1 atoms may be the key for the realization of the room temperature superconductivity. In order to improve the quality of the superconductivity caused by O1 vacancy, we suggest oxidation during the sample preparation or annealing in oxygen atmosphere as reported in the CuO nanostructures with the growth or post annealing in the different O atmosphere, the O vacancy decreased enormously[50]. And we believe that with such oxidation or annealing process under the oxygen atmosphere the quality of the superconductivity should be greatly improved.

In conclusion, our first-principles calculations shed light on the geometry and electronic structure of lead apatite with and without Cu-doped. Without the Cu doped, the lead apatite behaves as an insulator with the flat band straddling the Fermi level. However, the introduction of O1 vacancies eliminates these flat bands. We have identified the crucial role of the fully occupied p states of the 1/4 occupied O1 atoms in determining the quality of superconductivity in LK-99. The presence of Cu doping creates two one-dimensional channels, which may contribute to the realization of room-temperature superconductivity. However, the superconductivity is not observed in the experiments, possibly due to the missing of three out of four O1 atoms in the LK-99-O1 crystal. We propose that annealing in an oxygen atmosphere could be a potential solution to improve the quality of superconductivity. Further investigations are needed to fully comprehend the potential of Cu-doped lead apatite as a room-temperature superconductor.

ACKNOWLEDGEMENT

The authors thank Daqiang Gao, Peitao Liu for the fruitful discussion. This work is supported by the National Natural Science Foundation of China (Nos. 12174164, 91963201 and 11834005) and the 111 Project under Grant No. B2006 and Key R&D Project of Gansu Province (No. 22YF7WA014). The work was carried out at National Supercomputer Center in Tianjin, and the calculations were performed on Tianhe new generation supercomputer.

taokun@lzu.edu.cn

[1] P. Mangin and R. Kahn, "Superconductivity: An introduction" (Springer Cham, New York, 2016).

[2] J. Bardeen, L. N. Cooper, and J. R. Schrieffer, "Microscopic Theory of Superconductivity" Phys.Rev. 106, 162 (1957).

[3] J. G. Bednorz and K. A. Müller, Possible high Tc superconductivity in the Ba-La-Cu-O system, Zeitschrift für

Physik B Condensed Matter 64, 189 (1986).

[4] A. P. Drodzov, M. I. Eremets, I. A. Troyan, V. Ksenofontov, and S. I. Shylin, Conventional superconductivity at 203 kelvin at high pressures in the sulfur hydride system, Nature 525, 73 (2015).

[5] Y. Kamihara, T. Watanabe, M. Hirano, and H. Hosono, Iron-Based Layered Superconductor La[O1-xFx]FeAs (x = 0.05-0.12) with Tc = 26 K, Journal of the American Chemical Society 130, 3296 (2008).

[6] D. Li, K. Lee, B. Y. Wang, M. Osada, S. Crossley, H. R. Lee, Y. Cui, Y. Hikita, and H. Y. Hwang, Superconductivity in an infinite-layer nickelate, Nature 572, 624 (2019).

[7] Cao, Y.; Fatemi, V.; Demir, A.; Fang, S.; Tomarken, S. L.; Luo, J. Y.; Sanchez-Yamagishi, J. D.; Watanabe, K.; Taniguchi, T.; Kaxiras, E.; Ashoori, R. C.; Jarillo-Herrero, P. Correlated insulator behaviour at half-filling in magic-angle graphene superlattices, Nature, 556, 80 (2018).

[8] Cao, Y.; Fatemi, V.; Fang, S.; Watanabe, K.; Taniguchi, T.; Kaxiras, E.; Jarillo-Herrero, P.Unconventional superconductivity in magic-angle graphene superlattices, Nature 556, 43 (2018).

[9] N. Dasenbrock-Gammon, E. Snider, R. McBride, H. Pasan, D. Durkee, N. Khalvashi-Sutter, S. Munasinghe, S. E. Dissanayake, K. V. Lawler, A. Salamat, and R. P. Dias, Evidence of near-ambient superconductivity in a N-doped lutetium hydride, Nature 615, 244 (2023).

[10] X. Ming, Y.-J. Zhang, X. Zhu, Q. Li, C. He, Y. Liu, T. Huang, G. Liu, B. Zheng, H. Yang, J. Sun, X. Xi, and H.-H. Wen, Absence of near-ambient superconductivity in LuH2xNy, Nature, 620,72 (2023)

[11] X. Xing, C.Wang, L. Yu, J. Xu, C. Zhang, M. Zhang, S. Huang, X. Zhang, B. Yang, X. Chen, Y. Zhang, J. gang Guo, Z. Shi, Y. Ma, C. Chen, and X. Liu, "Observation of non-superconducting phase changes in LuH2xNy", Nat. Commun. 14, 5991 (2023)

[12] N. P. Salke, A. C. Mark, M. Ahart, and R. J. Hemley, "Evidence for Near Ambient Superconductivity in the Lu-N-H System",arXiv:2306.06301 (2023).

[13] D. Peng, Q. Zeng, F. Lan, Z. Xing, Y. Ding, and H. kwang Mao, "The near room-temperature upsurge of electrical resistivity in Lu-H-N is not superconductivity, but a metal-to-poor-conductor transition", Matter and Radiation at Extremes 8, 058401 (2023).

[14] S. Lee, J.-H. Kim, and Y.-W. Kwon, The First Room-Temperature Ambient-Pressure Superconductor, arXiv:2307.12008 (2023).

[15] S. Lee, J. Kim, H.-T. Kim, S. Im, S. An, and K. H. Auh, Superconductor $Pb_{10-x}Cu_x(PO_4)_6O$ showing levitation at room temperature and atmospheric pressure and mechanism, arXiv:2307.12037 (2023).

[16] Hao Wu, Li Yang, Bichen Xiao, Haixin Chang, Successful growth and room temperature ambient-pressure magnetic levitation of LK-99, arXiv:2308.01516 (2023).

[17] P. Abramian, A. Kuzanyan , V. Nikoghosyan , S. Teknowijoyo, and A. Gulian, Some remarks on possible superconductivity of composition $Pb_9CuP_6O$, arXiv: 2308.01723(2023).

[18] Sinéad M. Griffin, Origin of correlated isolated flat bands in copper-substituted lead phosphate apatite, arXiv:2307.16892v1(2023)

[19] D. Garisto, LK-99 isn't a superconductor -how science sleuths solved the mystery, Nature, 620, 705 (2023)

[20] D. Garisto, Claimed superconductor LK-99 is an online sensation -but replication efforts fall short, Nature, 620, 253 (2023)

[21] Hou, Q.; Wei, W.; Zhou, X.; Sun, Y.; Shi, Z. Observation of Zero Resistance above 100 K $Pb_{10-X}Cu_x(PO_4)_6O$. arXiv:2308.01192 (2023)

[22] Wu, H.; Yang, L.; Yu, J.; Zhang, G.; Xiao, B.; Chang, H. Observation of Abnormal Resistance-Temperature Behavior along with Diamagnetic Transition in $Pb_{10}XCu_x(PO_4)_6O$-Based Composite. arXiv:2308.05001 (2023).

[23] Wang, P., Liu, X., Ge, J. et al. Ferromagnetic and insulating behavior in both half magnetic levitation and non-levitation LK-99 like samples. Quantum Front 2, 10 (2023)

[24] Hou, Q.; Wei, W.; Zhou, X.; Wang, X.; Sun, Y.; Shi, Z. Current Percolation Model for the Special Resistivity Behavior Observed in Cu-Doped Apatite. arXiv:2308.05778 (2023)

[25] Liu, L.; Meng, Z.; Wang, X.; Chen, H.; Duan, Z.; Zhou, X.; Yan, H.; Qin, P.; Liu, Z. Semiconducting Transport in $Pb_{10-X}Cu_x(PO_4)_6O$ Sintered from $Pb_2SO_5$ and $Cu_3P$. Adv. Funct. Mater. 2308938 (2023)

[26] Kumar, K.; Karn, N. K.; Awana1, V. P. S. Synthesis of Possible Room Temperature Superconductor LK-99: P-

b9Cu(PO4)6O. Superconductor Science and Technology 36, 10 (2023)

[27] Cabezas-Escares, J.; Barrera, N. F.; Cardenas, C.; Munoz, F. Theoretical Insight on the LK-99 Material. arXiv:2308.01135 (2023).

[28] Lai, J.; Li, J.; Liu, P.; Sun, Y.; Chen, X.-Q. First-Principles Study on the Electronic Structure of Pb10-xCux(PO4)6O (X=0, 1). arXiv:2307.12008 (2023).

[29] Liu, R.; Guo, T.; Lu, J.; Ren, J.; Ma, T. Different Phase Leads to Different Transport Behavior in Pb9Cu(PO4)6O Compounds. arXiv:2308.08454 (2023).

[30] Yang, S.; Liu, G.; Zhong, Y. Ab Initio Investigations on the Electronic Properties and Stability of Cu-Substituted Lead Apatite (LK-99) Family with Different Doping Concentrations (X=0, 1, 2). arXiv:2308.13938 (2023).

[31] Korotin, D. M.; Novoselov, D. Y.; Shorikov, A. O.; Anisimov, V. I.; Oganov, A. R. Electronic Correlations in Promising Room-Temperature Superconductor Pb9Cu(PO4)6O: A DFT+DMFT Study. arXiv:2308.04301 (2023).

[32] Yue, C.; Christiansson, V.; Werner, P. Correlated Electronic Structure of Pb10-xCux(PO4)6O. arXiv:2308.04976 (2023).

[33] Zhang, J.; Li, H.; Yan, M.; Gao, M.; Ma, F.; Yan, X.W.; Xie, Z. Y. Structural, Electronic, Magnetic Properties of Cu-Doped Lead-Apatite Pb10-xCux(PO4)6O. arXiv:2308.04344 (2023).

[34] Jiang, Y.; Lee, S. B.; Herzog-Arbeitman, J.; Yu, J.; Feng, X.; Hu, H.; Calugaru, D.; Brodale, P. S.; Gormley, E. L.; Vergniory, M. G.; Felser, C.; Blanco-Canosa, S.; Hendon, C. H.; Schoop, L. M.; Bernevig, B. A. Pb9Cu(PO4)6(OH)2: Phonon Bands, Localized Flat Band Magnetism, Models, and Chemical Analysi. arXiv:2308.05143 (2023).

[35] Hao, L.; Fu, E. First-Principles Calculation on the Electronic Structures, Phonon Dynamics, and Electrical Conductivities of Pb10(PO4)6O and Pb9Cu(PO4)6O Compounds. J. Mater. Sci. Technol. 218, 173 (2024)

[36] Georgescu, A. B. Cu-Doped Pb10(PO4)6O, and V Doped SrTiO3 -a Tutorial on Electron-Crystal Lattice Coupling in Insulating Materials with Transition Metal Dopants. arXiv:2308.07295 (2023).

[37] Liu, J.; Yu, T.; Li, J.; Wang, J.; Lai, J.; Sun, Y.; Chen, X.-Q.; Liu, P. Symmetry Breaking Induced Insulating Electronic State in Pb9Cu(PO4)6O. arXiv:2308.11766 (2023).

[38] G. Baskaran, Broad Band Mott Localization is all you need for Hot Superconductivity: Atom Mott Insulator Theory for Cu-Pb Apatite, arXiv:2308.01307 (2023)

[39] Junwen Lai, Jiangxu Li, Peitao Liu, Yan Sun, and Xing-Qiu Chen, J. of Materials Science & Technology 171, 66 (2024).

[40] Liang Si and Karsten Held, Electronic structure of the putative room-temperature superconductor Pb9Cu(PO4)6O. Phys. Rev. B 108, L121110 (2023).

[41] G. Kresse and J. Hafner, Ab initio molecular dynamics for liquid metals, Phys. Rev. B 47, 558(R) (1993).

[42] G. Kresse and J. Furthmuller, Efficient iterative schemes for ab initio total-energy calculations using a plane-wave basis set, Phys. Rev. B 54, 11169 (1996).

[43] Krivovichev, S.V.; Burns, P.C, Crystal chemistry of Lead Oxide Phosphates: Crystal Structures of Pb4O(PO4)2, Pb4O8(PO4)2 and Pb10(PO4)6O. Z. Kristallogr. 218, 357, (2003)

[44] E. Gull and A. J. Millis, Numerical models come of age, Nature Physics 11, 808 (2015).

[45] H.-C. Jiang and T. P. Devereaux, Superconductivity in the doped Hubbard model and its interplay with next-nearest hopping t, Science 365, 1424 (2019).

[46] A. V. Krukau, O. A. Vydrov, A. F. Izmaylov, and G. E. Scuseria, J. Chem. Phys. 125, 224106 (2006).

[47] K. Tao, R. Chen, L. Yang, J. Gao, D. Xue, and C. Jia, The Cu induced ultraflat band in the room-temperature superconductor Pb10-xCux(PO4)6O4 (x = 0, 0.5) arxiv:2308.03218 (2023).

[48] P. Puphal, M. Y. P. Akbar, M. Hepting, E. Goering, M. Isobe, A. A. Nugroho, B. Keimer, Single crystal synthesis, structure, and magnetism of Pb10-xCux(PO4)6O, arXiv:2308.06256 (2023)

[49] Kaizhen Guo, Yuan Li, and Shuang Jia, Ferromagnetic half levitation of LK-99-like synthetic samples, Sci. China-Phys. Mech. Astron. 66, 107411 (2023).

[50] Daqiang Gao, Guijin Yang, Jinyun Li, Jing Zhang, Jinlin Zhang, and Desheng Xue, Room-Temperature Ferromagnetism of Flowerlike CuO Nanostructures, J. Phys. Chem. C. 114, 18347 (2010)